\documentclass[11pt]{article}

\usepackage{lineno}
\usepackage[export]{adjustbox}
\usepackage[margin=1in]{geometry}
\usepackage{amssymb,amsmath,latexsym}                        
\usepackage{graphicx}
\usepackage{cite}
\usepackage[table]{xcolor}%
\usepackage{float}

\usepackage{amsthm}
\usepackage{multicol}
\usepackage{authblk}
\usepackage{subfigure}
\usepackage{multirow} 
\let\emptyset\varnothing

\usepackage[hyperfootnotes=true]{hyperref}
\hypersetup{
 colorlinks=true,
 citecolor=blue,
 linkcolor=blue,
 urlcolor=blue}

\usepackage{adjustbox}
\usepackage{bbm}
\usepackage{mathtools}

\usepackage[titletoc]{appendix}
\date{\today}

\newcommand{\Zp}{$Z^\prime$}

\begin{document}

\title{The motivation and status of two-body \\ resonance decays after the LHC Run 2 and beyond}

\author[1]{Jeong Han Kim}
\author[1]{Kyoungchul Kong}
\author[2]{Benjamin Nachman}
\author[3]{Daniel Whiteson}
\affil[1]{\normalsize\it Department of Physics and Astronomy, University of Kansas, Lawrence, KS 66045, USA}
\affil[2]{\normalsize\it Physics Division, Lawrence Berkeley National Laboratory, Berkeley, CA 94704, USA}
\affil[3]{\normalsize\it Department of Physics and Astronomy, University of California, Irvine, CA 92697, USA}

\maketitle

\begin{abstract}
Searching for two-body resonance decays is a central component of the high energy physics energy frontier research program.  While many of the possibilities are covered when the two bodies are Standard Model (SM) particles, there are still significant gaps.  If one or both of the bodies are themselves non-SM particles, there is very little coverage from existing searches.  We review the status of two-body searches and motivate the need to search for the missing combinations.  It is likely that the search program of the future will be able to cover all possibilities with a combination of dedicated and model agnostic search approaches.
\end{abstract}

\section{Introduction}
\label{sec:intro}

One of the oldest and most fruitful methods for discovering new particles is to search for resonance structures in invariant mass spectra from the new particle decay products.  Most recently, this resulted in the discovery of the Higgs boson~\cite{Aad:2012tfa,Chatrchyan:2012xdj}, but has a long history from the direct observation of the $Z$ boson~\cite{Arnison:1983mk,Bagnaia:1983zx}, the discovery of the $\Upsilon$ (and thus $b$-quarks)~\cite{PhysRevLett.39.252}, the $J/\psi$ (and thus $c$-quarks)~\cite{PhysRevLett.33.1404,PhysRevLett.33.1406}, all the way to the $\rho$ meson~\cite{PhysRev.126.1858} and likely earlier.  This `bump hunting' continues to be a large component of the search program for the experiments at the Large Hadron Collider (LHC), with about a hundred searches in a multitude of final state configurations~\cite{atlasexoticstwiki,atlashdbstwiki,cmsexoticstwiki,cmsb2gtwiki,lhcbexoticstwiki}.  Unlike searches targeting more complex final states, for a given topology, two-body resonance searches are only sensitive to two parameters: the mass of the new particle and the production cross-section\footnote{There is also a mild dependence on the width, but this work will mostly consider narrow resonances where the width is small compared to the relevant experimental resolution.}.  As a result, these searches set powerful constraints on a variety of specific models of physics beyond the Standard Model (BSM).

Given that there have been no confirmed discoveries for new heavy particles since the discovery of the Higgs boson, it is critical to ensure that the complete landscape of two-body resonances is covered by the existing search program. The authors of Ref.~\cite{Craig:2016rqv} enumerated the possible scenarios and provided physics motivations for $A\rightarrow BC$, where $A$ is a BSM particle and $B$ and $C$ are SM particles.  One of our goals in this article is to provide a status update, given that the full Run 2 dataset has been collected and a number of searches have been performed since Ref.~\cite{Craig:2016rqv}.

While it is critical that dedicated searches targeting specific topologies continue to improve their scope and sensitivity, there is also a growing need for more model agnostic searches.  It may not be possible to have dedicated searches for every possible combination of SM particles for $B$ and $C$, and if either or both of these particles are themselves BSM particles, then the number of possibilities is endless.  Recently, there have been a variety of proposals to search for such scenarios in an automated manner using machine learning~\cite{Collins:2019jip,Collins:2018epr,Farina:2018fyg,Heimel:2018mkt,Cerri:2018anq,Roy:2019jae,Hajer:2018kqm,DeSimone:2018efk,DAgnolo:2018cun}.  Our second goal is to extend Ref.~\cite{Craig:2016rqv} to cases where $B$ and/or $C$ can be BSM particles to study the motivation and coverage of the complete two-body landscape.  This work may help focus on the application of the machine learning-based model agnostic searches.

This paper is organized as follows.  Section~\ref{sec:motivation} motivates two-body searches, for both the fully SM and mixed SM/BSM cases.  The status of existing experimental searches is presented in Section~\ref{sec:results}.  The paper ends with conclusions and outlook in Section~\ref{sec:conclusions}.

\section{Theory Motivation}
\label{sec:motivation}







Collider searches for resonances are well-motivated by their simplicity and a long history of discoveries. 
New resonances appear in many extensions of the SM and most of the experimental searches have followed the theoretical models, leading to 
a variety of searches for a pair of identical objects but rarely for non-identical pairs. 
However, there is no obvious compelling reason why one should focus only on identical pairs.
In fact, the diversity and simple structure of various resonances strongly motivate an experimental program which targets a broad scope and a systematic approach capable of theoretically unanticipated discoveries. 
Ref. \cite{Craig:2016rqv} proposed a systematic search program for 2-body resonances, which would consist of searches for resonances in all pairs of SM objects.  A majority of 2-body resonances have some indirect theoretical constraints but have received little experimental attention, leaving most of the landscape unexplored and a large potential for unanticipated discovery. It is interesting to note that the lack of these searches is not due to non-existence of theory models as there are models for all possible pairs.

In this article, we take a step further and generalize the final state of 2-body resonances to include BSM particles. 
We present our survey in various tables in this section. 
We begin with the main classification in Table \ref{tab:searches}, which contains 10 independent decay groups. 
Each row and column represent how $B$ and $C$ decay after the main decay process, $A \to B C$. 
The second cell in the first row ($B$) and the first column ($C$) represents a SM particle, while the other three cells represent a BSM particle. 
These three BSM cells are distinguished based on how they decay: 
$\text{BSM}\rightarrow \text{SM$_1$}\times\text{SM$_1$}$ (two similar kinds of SM particles), 
$\text{BSM}\rightarrow \text{SM$_1$}\times\text{SM$_2$}$ (two different kinds of SM particles), and 
$\text{BSM}\rightarrow \text{complex}$ (more complex final states). 
Our goal is not to provide a complete survey of all available theory models but to catalogue the set of possibilities, providing at least one motivating example for each final state\footnote{In nearly every case, there are multiple examples that have been well-studied in dedicated papers (we apologize for not citing your paper!).  This is particularly true for signatures that resemble all-hadronic diboson decays~\cite{Aad:2015owa} or contain di-photon resonances~\cite{Khachatryan:2016hje,Aaboud:2016tru} due to the excitement over (no longer) excesses reported by ATLAS and CMS.}.  

The left-upper corner of the table (denoted by Group I) reproduces a group of the standard 2-body decays, $A\rightarrow BC$, where $A$ is a BSM particle and $B$ and $C$ are SM particles, as covered in Ref. \cite{Craig:2016rqv}. 
In this subtable, the column and row are a list of SM particles and each entry corresponds to a mother particle, which would decay into one particle in the column and one particle in the row in the subtable. We show examples of theories that populate the entire landscape of 2-body resonances. 
$Z'$ and $W'$ denote additional gauge bosons, $\slash \hspace{-2.65mm}R$ represents $R$-parity violating supersymmetry (SUSY), $L^*, Q^*$ are excited leptons and quarks, respectively, and $T'$ and $B'$ are a vector-like top and bottom quarks, respectively.  $Z_{KK}$ denotes Kaluza-Klein excitation of SM $Z$.

We categorize the rest of Table \ref{tab:searches} in terms of nine additional subtables, which are denoted by Roman numerals II through X, and present each table in the sequential order. 
Note that generally we suppress electric charges of each SM particle and focus on the diversity of decay products, although we mention a few interesting examples of such kinds. Similarly we will not distinguish light jets from gluon and generically denote them as $j$ but occasionally we distinguish them for some interesting decays. 
We denote the bottom quark, and top quark by $b/\bar b$ and by $t /\bar t$, respectively.
The $V$ represents SM gauge bosons $Z$ and $W^\pm$ and $H$ is a SM Higgs boson. 
Throughout the manuscript, a primed particle $X^\prime$ represents a BSM particle, whose properties are similar to the corresponding SM particle $X$.

\begin{table}[t!]
\begin{adjustbox}{width=\columnwidth,center}

    \begin{tabular}{|cc||ccccccccc|c|c|c|}
    \hline
 \multicolumn{2}{|c||}{\multirow{2}{*}{$A\rightarrow BC$} } & \multicolumn{9}{c|}{$B=\text{SM}$} & $B=\text{BSM}$ & $B=\text{BSM}$& $B=\text{BSM}$ \\
   & & {$e$} &  {$\mu$} &  {$\tau$} &  {$q/g$} &  {$b$} &  {$t$} &  {$\gamma$} &  {$Z/W$} &  {$H$} & $\text{BSM}\rightarrow \text{SM$_1$}\times\text{SM$_1$}$ &  $\text{BSM}\rightarrow \text{SM$_1$}\times\text{SM$_2$}$ & $\text{BSM}\rightarrow \text{complex}$\\
       \hline
       \hline
      &{$e$} &$Z'$&$\slash \hspace{-2.7mm}R$&$\slash \hspace{-2.7mm}R$&$LQ$&$LQ$&$LQ$&$L^*$&$L^*$&$L^*$&&& \\
       &{$\mu$}& &$Z'$&$\slash \hspace{-2.7mm}R$&$LQ$&$LQ$&$LQ$&$L^*$&$L^*$&$L^*$&&& \\
       &{$\tau$}& & &$Z'$&$LQ$&$LQ$&$LQ$&$L^*$&$L^*$&$L^*$&&& \\
       & {$q/g$} & & &&$Z'$&$W'$&$T'$&$Q^*$&$Q^*$&$Q'$&&& \\
       \multirow{2}{*}{\rotatebox{90}{$C=\text{SM}$}} & $b$& &&&&$Z'$&$W'$&$Q^*$&$Q^*$&$B'$&Group II&Group III&Group IV \\
       &    {$t$} & &&&&&$Z'$&$Q^*$&$T'$&$T'$&(Table \ref{tab:searchII})& (Table \ref{tab:searchIII})& (Table \ref{tab:searchIV}) \\
       &    {$\gamma$} & & &&&&&$H$&$H$&$Z_{KK}$&&& \\
       &     {$Z/W$} & Group I  &  &&&&&&$H$&$H^\pm/A$&&& \\
       &     {$H$} &  & &&&&&&&$H$&&&\\
             \hline
             \parbox[t]{0mm}{\multirow{9}{*}{\rotatebox[origin=c]{90}{$C=\text{BSM}$}}} & \parbox[t]{0mm}{\multirow{9}{*}{\rotatebox[origin=c]{90}{$\text{BSM}\rightarrow \text{SM$_1$}\times\text{SM$_1$~ }$}}}    &&&&&&&&&&\multirow{9}{*}{\shortstack{Group V \\ (Table \ref{tab:searchV})}  } & \multirow{9}{*}{\shortstack{Group VI \\ (Table \ref{tab:searchVI})}} & \multirow{9}{*}{\shortstack{Group VII \\ (Table \ref{tab:searchVII})}} \\
		&&&&&&&&&&&   &   &   \\
		&&&&&&&&&&&   &   &   \\
		&&&&&&&&&&&   &   &   \\
		&&&&&&&&&&&   &   &   \\
		&&&&&&&&&&&   &   &   \\
		&&&&&&&&&&&   &   &   \\
		&&&&&&&&&&&   &   &   \\
		&&&&&&&&&&&   &   &   \\
                       \hline
                       \parbox[t]{0mm}{\multirow{9}{*}{\rotatebox[origin=c]{90}{$C=\text{BSM}$}}} & \parbox[t]{0mm}{\multirow{9}{*}{\rotatebox[origin=c]{90}{$\text{BSM}\rightarrow \text{SM$_1$}\times\text{SM$_2$~ }$}}}    &&&&&&&&&&& \multirow{9}{*}{\shortstack{Group VIII \\ (Table \ref{tab:searchVIII})}} & \multirow{9}{*}{\shortstack{ Group IX \\ (Table \ref{tab:searchIX})}} \\
		&&&&&&&&&&&   &   &   \\
		&&&&&&&&&&&   &   &   \\
		&&&&&&&&&&&   &   &   \\
		&&&&&&&&&&&   &   &   \\
		&&&&&&&&&&&   &   &   \\
		&&&&&&&&&&&   &   &   \\
		&&&&&&&&&&&   &   &   \\
		&&&&&&&&&&&   &   &   \\
                       \hline
                       \parbox[t]{0mm}{\multirow{9}{*}{\rotatebox[origin=c]{90}{$C=\text{BSM}$}}} & \parbox[t]{0mm}{\multirow{9}{*}{\rotatebox[origin=c]{90}{$\text{BSM}\rightarrow \text{complex~~}$}}}    &&&&&&&&&&&  & \multirow{9}{*}{\shortstack{ Group X \\ (Tables \ref{tab:searchX} and \ref{tab:searchX2})}} \\
		&&&&&&&&&&&   &   &   \\
		&&&&&&&&&&&   &   &   \\
		&&&&&&&&&&&   &   &   \\
		&&&&&&&&&&&   &   &   \\
		&&&&&&&&&&&   &   &   \\
		&&&&&&&&&&&   &   &   \\
		&&&&&&&&&&&   &   &   \\
		&&&&&&&&&&&   &   &   \\
                       \hline
    \end{tabular}

  \end{adjustbox}
  \caption{\label{tab:searches} Top-level organization of BSM particle $A$ by its two-body decays into $B$ and $C$, showing examples of theoretical motivations for each case.  $Z'$ and $W'$ denote additional gauge bosons, $\slash \hspace{-2.7mm}R$ represents $R$-parity violating SUSY, $L^*, Q^*$ are excited leptons and quarks, respectively, and $T'$ and $B'$ are a vector-like top and bottom quarks, respectively.  The symbol $Z_{KK}$ denotes Kaluza-Klein excitation of SM $Z$.  The SM case in the upper left box is reproduced from Ref.~\cite{Craig:2016rqv}.}
\end{table}

Table \ref{tab:searchII} shows example for $A \to B C$, where $A$ and $B$ are BSM particles and $C$ is a SM particle, which is the Group II in Table \ref{tab:searches}. We consider two similar SM particles in the $B$ decays. 
For example, the $jj$ denotes $B$ decays to two quarks ($q\bar q$, $q \bar q^\prime$ or $q q$), while $\ell\ell$ includes both two opposite-charged leptons ($\ell^+\ell^-$) and the same-sign charged leptons ($\ell^+ \ell^+$ and $\ell^- \ell^-$)
and the $VV$ includes the $B$ decays to $gg$, $\gamma\gamma$, $\gamma Z$, $ZZ$, $WW$, or $ZW$.
The $H$ is the observed Higgs boson, $H^{\prime \prime}$ and $H^\prime$ are heavy scalars, $A$ is a new pseudo scalar, and $ H^{++}$ denotes a doubly-charged scalar particle.
The $Q'$ represents a generic vector-like quark. 
$X_{5/3}$ and $\pi_6^{4/3}$ represent a vector-like quark with electric charge $5/3$ and a color-sextet scalar with electric charge $4/3$, respectively. 
Since we consider two similar SM particles, many such examples are either a $Z^\prime$/$W^\prime$ or a neutral-heavy scalar.  

\begin{table}[t!]
\centering
\renewcommand{\arraystretch}{1.1}
\setlength\tabcolsep{3pt}
\begin{tabular}{|c|c|| c | c| c| c| }
\hline 
\multicolumn{2}{|c||}{\multirow{2}{*}{$A\rightarrow BC$}} & \multicolumn{4}{c|}{$B=\text{BSM}$} \\
\cline{3-6} 
\multicolumn{2}{|c||}{}							& 	$\ell \ell$	&	$j j$			& 	$VV $	& $HH$      \\
\hline 
\hline 
&$\ell (e, \mu, \tau)$	&  $L^\prime\to \ell Z^\prime$ & $L^\prime\to \ell Z^\prime$, $N^\prime\to \ell W^\prime$ & $L^\prime\to \ell Z^\prime$, $N^\prime\to \ell W^\prime$ & $L^\prime\to \ell H^\prime$	  \\
\cline{2-6} 
\multirow{5}{*}{\rotatebox{90}{\hspace*{-0.7cm}$C=\text{SM}$}}			&	&	$Q^\prime\to j Z^\prime$, & $Q^\prime\to j Z^\prime$, & $Q^\prime\to j Z^\prime$, $Q^\prime\to j W^\prime$, &   	\\
&$j (b, t, q)$                 &$Q^\prime\to j H^\prime$,	 & $Q^\prime\to j H^\prime$, & $Q^\prime\to j H^\prime$,   &  $Q^\prime\to j H^\prime$	\\
&				& $X_{5/3}\to  b H^{++}$ 	 & $X_{5/3}\to \bar b \pi_6^{4/3}$   & $X_{5/3}\to  b H^{++}$  &  	\\
\cline{2-6} 
&\multirow{2}{*}{$V (W, \gamma, Z)$}	& \multirow{2}{*}{$W^\prime \to W Z^\prime$} 	& $W^\prime\to W Z^\prime$,   & $W^\prime\to W Z^\prime$, 	& $Z^\prime \to Z H^\prime$,     \\
 &                           	                         & 									& $Z^\prime\to W W^\prime$ 	& $Z^\prime\to W W^\prime$ 	&  $Z^\prime \to \gamma H^\prime$     \\
\cline{2-6}
&\multirow{2}{*}{$H$}				& $A \to H Z^\prime$,  	&$A \to H Z^\prime$,  &$A \to H Z^\prime$,  & $H^{\prime \prime} \to H H^\prime$    	\\
&                          	& $H^{\prime \prime} \to H H^\prime$  &  $H^{\prime \prime} \to H H^\prime$    &   $H^{\prime \prime} \to H H^\prime$ &  	   \\
\hline
\end{tabular}
\caption{\label{tab:searchII} Example theoretical models for two-body decay of a BSM particle $A$ into a BSM particle $B$ and an SM particle $C$, where the $B$ particle subsequently decays to two similar SM particles (Group II in Table \ref{tab:searches}).
The $jj$ denotes $B$ decays to $q\bar q$/$q \bar q^\prime$/$qq$,
and the $VV$ includes the $B$ decays to $gg$, $\gamma\gamma$, $\gamma Z$, $ZZ$, $WW$, or $ZW$.
The $H$ is the observed Higgs boson, $H^{\prime \prime}$ and $H^\prime$ are heavy scalars, $A$ is a new pseudo scalar, and $ H^{++}$ denotes a doubly-charged scalar particle.
The $Q'$ represents a generic vector-like quark. In particular, an exotic vector-like quark with electric charge $5/3$ is denoted as $X_{5/3}$.
The $\pi_6^{4/3}$ is a color-sextet scalar with electric charge $4/3$.}
\end{table}

\begin{figure}[t!]
\centering
\includegraphics[width=0.95\textwidth]{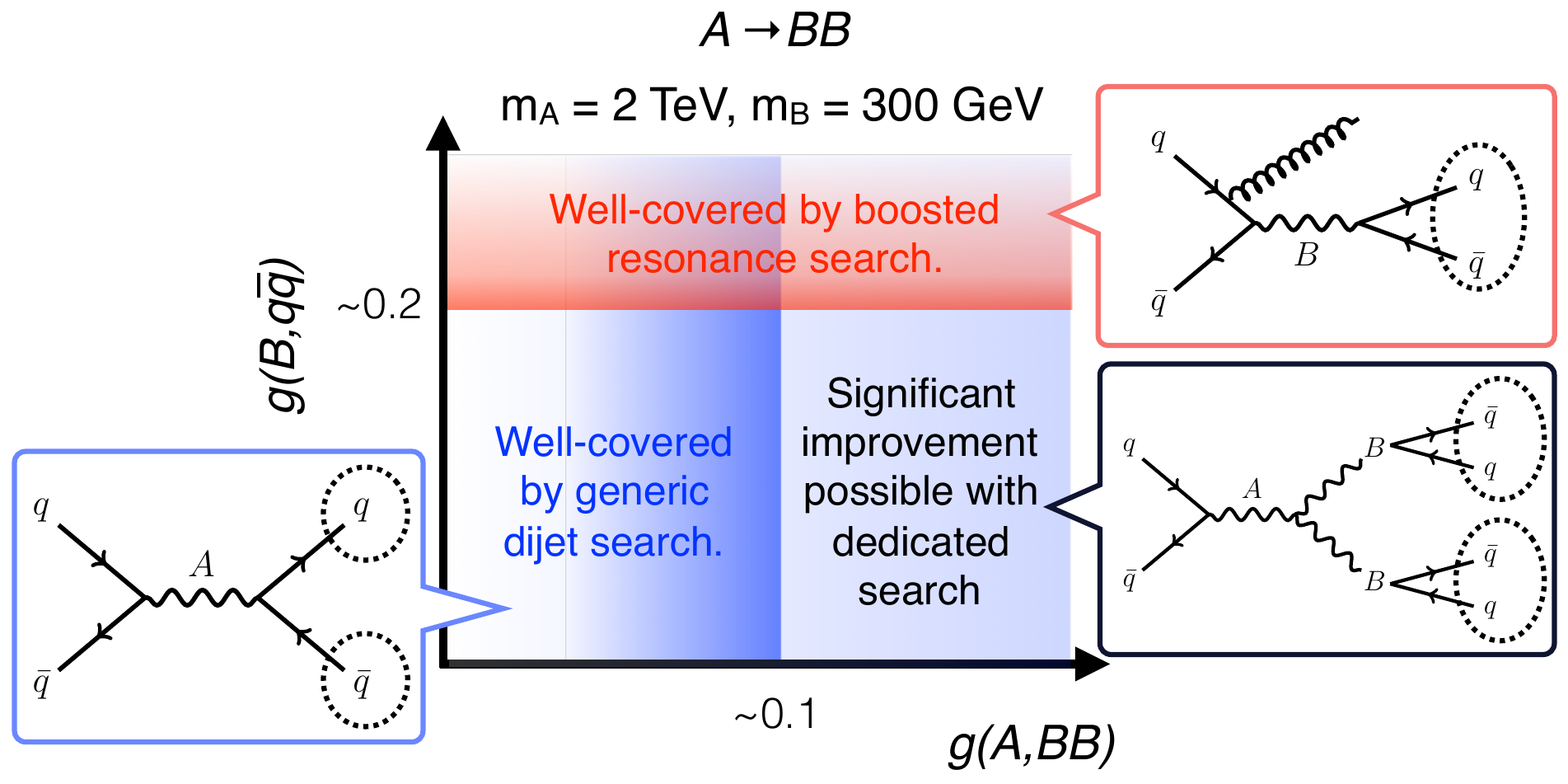}
\caption{\label{fig:diagram} An illustration of the complementarity of the search for $A$ (inclusive dijet resonance search) and the search for $B$ (boosted resonance search).  Dotted circles indicate hadronic activity that will likely be mostly captured by one (potentially large-radius) jet.  When $m_A=2$ TeV and $m_B=300$ GeV, the inclusive dijet search likely has reduced sensitivity to $A\rightarrow BB$ because the $B$ decay products are not well-contained inside a single small-radius jet.  Therefore, when $g(A,BB)\gtrsim g(A,q\bar{q})\sim 0.1$, gains are possible for a dedicated search.}
\end{figure}

It is worth noting that when $B$ or $C$ are BSM particles, searches for $A$ are complemented by searches for the $B$ or $C$ particle directly.  These approaches are complementary because searches for $A \to B C$ are sensitive to the coupling between the $A$ and $B/C$ while direct searches for $B/C$ are sensitive to the coupling between $B/C$ and the SM decay products.  It is possible that one of these couplings could be sufficiently smaller than the other to render direct searches in one mode insensitive and therefore both search strategies are useful.  Figure~\ref{fig:diagram} illustrates the complementarity of direct and indirect searches in the case that $B=C$ and $B\rightarrow q\bar{q}$.  The three relevant couplings are between the $A$ particle and quarks ($g(A,q\bar{q})$), between the $B$ particle ($g(A,BB)$) and between the $B$ particle and quarks ($g(B,q\bar{q})$).  When $m_B\ll m_A$, so that the $B$ decay products are contained inside a single jet, the inclusive dijet search sets strong limits on $A$ production.   These limits would be significantly weaker when $m_B$ is not sufficiently small for its decay products to be contained inside an $R=0.4$ jet, which is the jet radius used by both the ATLAS and CMS inclusive dijet searches.  For $m_A=2$ TeV, the current limit on $g(A,q\bar{q})$ is about 0.1~\cite{ATLAS-CONF-2019-007,CMS:2018wxx}.  For moderate (not contained) $m_B$, this means that there is strong sensitivity up to $g(A,BB)\sim g(A,q\bar{q})$.  For larger $g(A,BB)$, there would be stronger sensitivity from a direct search that targets the full $A\rightarrow BB$ topology, e.g. a search for two jets with substructure and not just a search for two generic quark/gluon jets.  The coupling $g(A,q\bar{q})\sim 0.1$ at $m_A\sim 2$ TeV corresponds to a cross section limit of about 0.1 pb.   The direct search for $B$ sets limits of about 1 nb at $m_B=300$ GeV, which corresponds to $g(B,q\bar{q})\approx 0.2$~\cite{Aaboud:2019zxd,Sirunyan:2017nvi}.   Therefore, the direct $B$ search is not sensitive to the $B$'s produced from $A$ production.  However, the direct search for $B$ can be competitive when $g(A,q\bar{q})$ is small.  In particular, if $g(A,q\bar{q}) < 0.1$, then the direct search for $A$ is insensitive, but if $g(B,q\bar{q}) > 0.2$, then the $B$ search is sensitive. In general, this is also true for other final states and we expect significant improvement possible with a dedicated search for $A \to B C \to (SM SM)(SM SM)$, especially in the parameter space where $g(A, B C) \gtrsim g(A, SM SM)$ and $g(B, SM SM)$ / $g(C, SM SM)$ is not too large.

In next group (Group III in Table \ref{tab:searches}), we consider the case where $B$ decays to two different types of SM particles. 
Unlike Table \ref{tab:searchII} where many examples are either a $Z^\prime$ or a neutral-heavy scalar, in this category of Table \ref{tab:searchIII}, many examples of $B$ are either a vector-like quark, a charged scalar or a $W^\prime$, since we consider two different particles. Once the spin of $B$ is fixed, we can easily find the spin nature of $A$, for a given SM particle for $C$. 
Here the $LQ$ is a leptoquark carrying both baryon and  lepton numbers, and the $N$ is a right-handed heavy neutrino.
Exotic vector-like quarks with electric charges $-7/3$, $-4/3$, $5/3$, and $8/3$ are denoted as $X_{-7/3}$, $X_{-4/3}$, $X_{5/3}$, and $X_{8/3}$ respectively. The $H^+$ and $H^-$ are new charged scalar particles.

\begin{table}[h!]
\renewcommand{\arraystretch}{1.05}
\centering
\scalebox{0.96}{
\begin{tabular}{|c|c|| c | c| c|c|}
\hline 
\multicolumn{2}{|c||}{\multirow{3}{*}{$A\rightarrow BC$}}  & \multicolumn{4}{c|}{$B=\text{BSM}$} \\
\cline{3-6} 
\multicolumn{2}{|c||}{}  &  $tZ$, $tH$, $Wb$, 		& $bZ, bH, Wt, $ 		& \multirow{2}{*}{$\ell Z$, $\ell\gamma$, or $\ell H$}  	& $\gamma W$, $ZW$, $HW,$\\
\multicolumn{2}{|c||}{}  &   $tg$, or $t\gamma$  		&  $bg$, or $b\gamma$	&    							&  or $t b$  \\
\hline 
\hline 
&$\ell $	&  	$LQ \to \ell T^\prime$	& $LQ \to \ell B^\prime$	& $H^\prime \to \ell L^\prime$	& $N \to \ell W^\prime$, $N\to \ell H^+ $\\
\cline{2-6} 
\multirow{5}{*}{\rotatebox{90}{\hspace*{-0.5cm}$C=\text{SM}$}}	&     & $W^\prime \to b T^\prime$,	& $W^\prime \to t B^\prime$, 	&  $LQ \to j L^\prime$  & $B^\prime\to t W^\prime$, $T^\prime\to b H^+$,  \\
&$j (b, t, q)$    & $Z^\prime \to t T^\prime$  	&  $Z^\prime \to b B^\prime$	&  &  $X_{5/3}\to t W^\prime$, $ B^\prime\to t H^-$,  \\
	        &     & 	&  &  &    $X_{5/3}\to  t H^+$  \\
\cline{2-6}  
&\multirow{2}{*}{ $V (W, \gamma, Z)$	} & $B^\prime\to W T^\prime$,	& $T^\prime\to W B^\prime$, 	&  $L^{\prime\prime} \to Z L^\prime$  &  $H^+\to \gamma W^\prime$,	 \\
&	                                                     & $X_{-7/3} \to W X_{-4/3}$   & $X_{8/3}\to W X_{5/3}$&     &	$Z^\prime \to W H^+$   \\
\cline{2-6} 
&\multirow{2}{*}{$H$}				& \multirow{2}{*}{$T^{\prime\prime}\to H T^\prime$}  	& \multirow{2}{*}{$B^{\prime\prime}\to H B^\prime$} & \multirow{2}{*}{$L^{\prime\prime} \to H L^\prime$}  & $W^{\prime\prime} \to H W^\prime$ \\
&				&   	&  &   &    $H^{+ \prime} \to H H^+$ \\
\hline
\end{tabular}}
\caption{\label{tab:searchIII} Example (Group III in Table \ref{tab:searches}) for $A \to B C$, where $A$ and $B$ are BSM particles and $C$ is a SM particle. 
The $B$ decays to two different SM particles are considered.
The $LQ$ is a leptoquark carrying both baryon and  lepton numbers, and the $N$ is a right-handed heavy neutrino.
Exotic vector-like quarks with electric charges $-7/3$, $-4/3$, $5/3$, and $8/3$ are denoted as $X_{-7/3}$, $X_{-4/3}$, $X_{5/3}$, and $X_{8/3}$ respectively.
The $H^+$ and $H^-$ are new charged scalar particles. 
}
\end{table}


The next example (Group IV) shown in Table \ref{tab:searchIV} is similar to the previous case (Group III), but we consider the case where $B$ decays in a more complicated way.
Many examples that we present are due to 3-body decay or decays through $R$-parity violating interaction (RPV) in supersymmetry.
For instance, a right-handed heavy neutrino $N$ could decay to $N \to W^{\prime (*)} \ell \to j j \ell $ via an off-shell $W^\prime$. 
Similarly, a generic vector-like quark $Q^\prime$ could decay to $Q^\prime \to W^{\prime (*)} j \to j j j$. 
In Table \ref{tab:searchIV}, the $\tilde t$ denotes a stop, and the $\tilde g$ is a gluino which decays through RPV couplings \cite{Barbier:2004ez, Evans:2012bf}. 
The $\tilde{W}^{0}$ and $\tilde{W}^{\pm}$ are neutral and charged winos respectively.
The neutral and charged Higgsinos, $\tilde{H}^{0}$ and $\tilde{H}^{\pm}$, can decay to $\tilde{H}^{\pm/0} \to j \tilde q^* \to jjj$ through RPV couplings.

\begin{table}[t!]
\centering
\renewcommand{\arraystretch}{1.1}
\setlength\tabcolsep{3pt}
\begin{tabular}{|c|c|| c | c| c| }
\hline 
\multicolumn{2}{|c||}{\multirow{3}{*}{$A\rightarrow BC$}} & \multicolumn{3}{c|}{$B=\text{BSM}$} \\
\cline{3-5} 
\multicolumn{2}{|c||}{} & 	$N (\to jj \ell )$,	$\tilde g (\to jj \ell )$, & $Q^\prime (\to jjj)$, $\tilde{H}^{\pm/0} (\to jjj)$,  & $LQ (\to \ell j)$, $\tilde q (\to  \ell j)$,  \\
\multicolumn{2}{|c||}{} & or $\tilde{W}^{\pm/0} (\to jj \ell )$	& or $\tilde g (\to jjj)$	& or $\tilde q (\to jj)$				\\
\hline 
\hline 
\multirow{5}{*}{\rotatebox{90}{\hspace*{-1cm}$C=\text{SM}$}}&$\ell (e, \mu, \tau)$	&	$W^\prime \to \ell  N$		&	$LQ \to \ell Q^\prime$ &			$Q^\prime \to \ell LQ$	\\
\cline{2-5} 
&                    		&	$LQ\to j N$,				& 	$\tilde t \to t \tilde g$,  &	\multirow{3}{*}{  	$L^\prime \to q LQ $}  \\
&$j (b, t, q)$                &	$\tilde t \to t \tilde g$,	$\tilde t \to t \tilde{W}^{0}$, & $\tilde q \to \tilde{H}^{\pm/0}  q$ 	 	& 	 \\
&		                &	$\tilde t \to b \tilde{W}^{\pm}$	& 	 	& 	 \\
\cline{2-5} 
&\multirow{2}{*}{ $V (W, \gamma, Z)$} &	\multirow{2}{*}{ $L^\prime \to W N$} 	   &	         & $\tilde q^\prime \to W \tilde q$	  \\
&                          	                           &	 		                     &  	 &	 $\tilde q^\prime \to Z \tilde q$	  \\
\cline{2-5} 
&$H$				&	$N^{\prime} \to H N$	&  &    $LQ^\prime \to H LQ$ \\
\hline
\end{tabular}
\caption{\label{tab:searchIV} Example (Group IV in Table \ref{tab:searches}) for $A \to B C$, where $A$ and $B$ are BSM particles and $C$ is a SM particle. 
The $B$ decays to more complex final states are considered.
The $L^\prime$ is a vector-like lepton, and $N$ denotes a right-handed heavy neutrino which can decay to $N \to W^{\prime (*)} \ell \to j j \ell $.
The $Q^\prime$ represents a generic vector-like quark decaying to $Q^\prime \to W^{\prime (*)} j \to j j j$. 
The $\tilde t$ denotes a stop, and the $\tilde g$ is a gluino which decays through RPV couplings \cite{Barbier:2004ez, Evans:2012bf}. 
The $\tilde{W}^{0}$ and $\tilde{W}^{\pm}$ are neutral and charged winos respectively.
The neutral and charged Higgsinos, $\tilde{H}^{0}$ and $\tilde{H}^{\pm}$, can decay to $\tilde{H}^{\pm/0} \to j \tilde q^* \to jjj$ through RPV couplings.
}
\end{table}


The next group (Group V in Table \ref{tab:searches}) presented in Table \ref{tab:searchV} is the first example of $A \to B C$ decay, where $A$, $B$ and $C$ are all BSM particles. 
We consider that each of $B$ and $C$ decays to similar kinds of SM particles. 
As discussed in Table  \ref{tab:searchII}, the $jj$ denotes two quark-system of all possible flavor combinations.
Here the $VV$ includes $\gamma \gamma$, $\gamma Z$, $ZZ$, $ZW$, $\gamma W$, or $WW$.
We abbreviate the decays $G^\prime_{\Theta \phi_I} \equiv G^\prime \to\Theta \phi_I$,  $G^\prime_{\Theta\Theta} \equiv G^\prime \to\Theta\Theta$, $\Theta \equiv \Theta \to G^\prime G^\prime$, and $\phi_I \equiv \phi_I \to \Theta G^\prime$ where $G^\prime$, $\Theta$, and $\phi_I$ denote a coloron, color-octet scalar, and a singlet scalar, respectively \cite{Bai:2018jsr, Bai:2018wnt}.  The extended Two-Higgs Doublet Model with a real singlet (2HDMS) \cite{Branco:2011iw, Chen:2013jvg} allows for the decay of a CP-even heavy scalar into light scalars, abbreviated as $2H \equiv H^{\prime \prime} \to H^{\prime} H^{\prime}$.  We also consider heavy $Z$ boson decays $Z^\prime \equiv Z^\prime\to W^{\prime +}W^{\prime -} \,, H^+ H^- \,,W^{\prime \pm} H^{\mp}, \text{or } H^{++} H^{--}$ where $H^\pm$ and $H^{\pm \pm}$ are singly- and doubly-charged scalars. 

\begin{table}[h!]
\centering
\hspace*{-0.0cm}
\scalebox{0.96}{
\begin{tabular}{|c|c|| c |c|c|c|c|}
\hline 
\multicolumn{2}{|c||}{\multirow{2}{*}{$A\rightarrow BC$}} & \multicolumn{5}{c|}{$B=\text{BSM}$} \\
\cline{3-7} 
\multicolumn{2}{|c||}{} 	& $jj$  & $gg$ & $VV$   & $\ell \ell$  & $HH$   \\
\hline 
\hline 
\multirow{5}{*}{\rotatebox{90}{$C=\text{BSM}$}}	&$jj$ & $\Theta$, \Zp  & $\phi_I$ & 2H, \Zp   & 2H  & 2H     \\
\cline{2-7} 
&$gg$ 	&   & $G^\prime_{\Theta\Theta}$ & 2H, $G^\prime_{\Theta \phi_I}$  &  2H &2H  \\
\cline{2-7} 
&$VV$  &   &  & 2H, \Zp   &  2H & 2H  \\
\cline{2-7} 
&$\ell \ell$ &   &   &   &  2H & 2H  \\
\cline{2-7} 
&$HH$ &   &   &      &     &  2H  \\
\hline 
\end{tabular}}
\caption{\label{tab:searchV} Example (Group V in Table \ref{tab:searches}) for $A \to B C$, where $A$, $B$ and $C$ are BSM particles. 
Each of $B$ and $C$ decays to similar kinds of SM particles.
The $jj$ denotes a diquark with all possible flavor combinations.
Here the $VV$ includes $\gamma \gamma$, $\gamma Z$, $ZZ$, $ZW$, $\gamma W$, or $WW$.
We abbreviate the decays $G^\prime_{\Theta \phi_I} \equiv G^\prime \to\Theta \phi_I$,  $G^\prime_{\Theta\Theta} \equiv G^\prime \to\Theta\Theta$, $\Theta \equiv \Theta \to G^\prime G^\prime$, and $\phi_I \equiv \phi_I \to \Theta G^\prime$ where $G^\prime$, $\Theta$, and $\phi_I$ denote a coloron, color-octet scalar, and a singlet scalar respectively \cite{Bai:2018jsr, Bai:2018wnt}. Extended Two-Higgs Doublet Model with a real singlet (2HDMS) \cite{Branco:2011iw, Chen:2013jvg} allow for the decay of a CP-even heavy scalar into light scalars, abbreviated as $2H \equiv H^{\prime \prime} \to H^{\prime} H^{\prime}$.  We also consider heavy $Z$ boson decays $Z^\prime \equiv Z^\prime\to W^{\prime +}W^{\prime -} \,, H^+ H^- \,,W^{\prime \pm} H^{\mp}, \text{or } H^{++} H^{--}$ where $H^\pm$ and $H^{\pm \pm}$ are singly- and doubly-charged scalars. }
\end{table}


The next two Groups, VI and VII, in Tables \ref{tab:searchVI} and \ref{tab:searchVII} are similar to Groups III and IV, respectively. In both cases, $C$ decays to similar kind SM particles, while $B$ decays to two different kinds (Group VI, Table \ref{tab:searchV} ) or more complex final state (Group VII, Table \ref{tab:searchVI}).

\begin{table}[t!]
\centering
\begin{tabular}{|c|c|| c | c| c| c|}
\hline 
\multicolumn{2}{|c||}{\multirow{3}{*}{$A\rightarrow BC$}}& \multicolumn{4}{c|}{$B=\text{BSM}$} \\
\cline{3-6} 
\multicolumn{2}{|c||}{} & 	$tZ$, $tH$, $Wb$, 					&	$bZ, bH, Wt, $ 		    & 	\multirow{2}{*}{$\ell Z$, $\ell\gamma$, or $\ell H$} 	& \multirow{2}{*}{$HW$}    \\
\multicolumn{2}{|c||}{}                                                      &      $tg$, or $t\gamma$                                    &       $bg$, or $b\gamma$    &                                                  &          \\
\hline 
\hline 
\multirow{3}{*}{\rotatebox{90}{\hspace*{-0.5cm}$C=\text{BSM}$}}&\multirow{2}{*}{ $jj$ }       	&	\multirow{2}{*}{$T^{\prime\prime}\to Z^\prime T^\prime$}    & \multirow{2}{*}{$B^{\prime\prime}\to Z^\prime B^\prime$	}  & \multirow{2}{*}{$L^{\prime\prime}\to Z^\prime L^\prime$}  & $ Z^\prime \to H^{\pm} W^{\prime \mp} $,  \\
                                          &                                                                           &                                           &                             &                                                                   & $ Z^\prime \to  W^{\prime +}W^{\prime -}  $     \\
\cline{2-6}
&$VV$	&	$T^{\prime\prime}\to S^\prime T^\prime$    & $B^{\prime\prime}\to S^\prime B^\prime$	& $L^{\prime\prime}\to S^\prime L^\prime$  &   \\
\cline{2-6}
&$HH$	&	$T^{\prime\prime}\to S^\prime T^\prime$    & $B^{\prime\prime}\to S^\prime B^\prime$	& $L^{\prime\prime}\to S^\prime L^\prime$   & \\
\hline
\end{tabular}
\caption{\label{tab:searchVI} Example (Group VI in Table \ref{tab:searches}) for $A \to B C$, where $A$, $B$ and $C$ are BSM particles.
The $C$ decays to two similar SM particles, while the $B$ decays to two different kinds of SM particles.
The $jj$ denotes a diquark with all possible flavor combinations.
The $VV$ includes $gg, \gamma \gamma$, $\gamma Z$, $ZZ$, or $WW$.
}
\end{table}



\begin{table}[t!]
\centering
\renewcommand{\arraystretch}{1.1}
\setlength\tabcolsep{3pt}
\begin{tabular}{|c|c|| c | c| c| c|}
\hline 
\multicolumn{2}{|c||}{\multirow{2}{*}{$A\rightarrow BC$}}& \multicolumn{4}{c|}{$B=\text{BSM}$} \\
\cline{3-6} 
\multicolumn{2}{|c||}{} 		& 	$ j j \ell$						&	$\ell \ell \ell$	  &  $tWW$ & $j j j$ \\
\hline 
\hline 
\multirow{3}{*}{\rotatebox{90}{\hspace*{-0.2cm}$C=\text{BSM}$}}&\multirow{3}{*}{ $jj$, $VV$, or $HH$ }	& $N^{\prime} \to Z^\prime N$,  & $L^{\prime\prime} \to Z^\prime L^\prime$, &  $ X_{8/3}^{\prime}   \to Z^{\prime} X_{8/3}   $ , & \multirow{3}{*}{$\tilde q \to \tilde{H}^+ \tilde q^\prime$ }    \\
        &                                                  & $N^{\prime} \to H^\prime N$,  & $L^{\prime\prime} \to H^\prime L^\prime$  &  $ X_{8/3}^{\prime}   \to H^{\prime} X_{8/3}   $  &   \\
        &                                                   &  $L^\prime \to W^\prime N$  &	                                                            &                                                                             &	 \\
\hline
\end{tabular}
\caption{\label{tab:searchVII} Example (Group VII in Table \ref{tab:searches}) for $A \to B C$, where $A$, $B$ and $C$ are BSM particles.
The $C$ decays to two similar SM particles, while the $B$ decays to more complex final states.
The $jj$ denotes a diquark with all possible flavor combinations, and $VV$ includes $ \gamma \gamma$, $\gamma Z$, $ZZ$, or $WW$.
The $N$ denotes a right-handed heavy neutrino which decays to $N \to W^{\prime (*)} \ell \to j j \ell $.
The $L^\prime$ represents a generic vector-like lepton decaying to $L^\prime \to Z^{\prime (*)} \ell \to \ell \ell \ell$. 
}
\end{table}


In Table \ref{tab:searchVIII}, we present examples for $A \to B C$, where $A$, $B$, and $C$ are BSM particles and both $B$ and $C$ decay to different kinds of SM particles (Group VIII in Table \ref{tab:searches}). 
As mentioned for Table \ref{tab:searchIII}, many examples of two different SM decay products are decays of vector-like fermions.
Therefore an obvious example for Group VIII would be the resonant production of two vector-like fermions (via either a new gauge boson or a new scalar).

\begin{table}[t!]
\centering
\renewcommand{\arraystretch}{1.1}
\begin{tabular}{|c|c||c|c|c|}
\hline 
\multicolumn{2}{|c||}{\multirow{2}{*}{$A\rightarrow BC$}} & \multicolumn{3}{c|}{$B=\text{BSM}$} \\
\cline{3-5} 
\multicolumn{2}{|c||}{} & { $Wb$,  $tZ$,	 $tH$,	  $tg$,  or $t \gamma$  }  & { $Wt$, $bZ$, $bH$, $bg$,	or $b \gamma$ } &  { $\ell Z$ or $\ell \gamma$} \\
\hline \hline
\multirow{5}{*}{\rotatebox{90}{\hspace*{-0.5cm}$C=\text{BSM}$}}&$Wb$, $tZ$,     &  { $Z^\prime \to T^\prime \bar{T^\prime}$,  } 	&     {  $W^\prime \to T^\prime \bar{B^\prime}$,  }      &       { \multirow{2}{*}{ $LQ \to T^\prime \bar{L^\prime}$ }  }          \\
&$tH$, $tg$, or $t \gamma$      &   { $ H^\prime \to T^\prime \bar{T^\prime}$    } 	&     {  $ H^\pm \to T^\prime \bar{B^\prime}$  }   &  {  }          \\
\cline{2-5}
&$Wt$, $bZ$,	&        {   } &     {  $Z^\prime \to B^\prime \bar{B^\prime}$,  }                   &       { \multirow{2}{*}{ $LQ \to B^\prime \bar{L^\prime}$}  }         \\
&$bH$, $bg$,or $b\gamma$	&       {   } &     {$H^\prime \to B^\prime \bar{B^\prime}$  }                   &       {   }          \\
\cline{2-5}
&\multirow{2}{*}{$\ell Z$ or $\ell \gamma$} 	&        {   } &     {   }                   &       { $Z^\prime \to L^\prime \bar{L^\prime}$  }          \\
&                                                                  &        {   } &     {   }                   &       {$S \to L^\prime \bar{L^\prime}$   }          \\
\hline
\end{tabular}
\caption{\label{tab:searchVIII} Example (Group VIII in Table \ref{tab:searches}) for $A \to B C$, where $A$, $B$, and $C$ are BSM particles.
Each of $B$ and $C$ decays to different kinds of SM particles.
This table shows the resonant productions (via either a new gauge boson or a new scalar) of new fermions. }
\end{table}


The last two Groups (IX and X) involve more complex decays.
In Table \ref{tab:searchIX}, only $B$ follows the complex decays, while in Table \ref{tab:searchX} both $B$ and $C$ give the complex final states.
We consider 3-body decays of a right-handed heavy neutrino or a vector-like quark for such examples.

\begin{table}[h!]
\centering
\renewcommand{\arraystretch}{1.1}
\begin{tabular}{|c|c|| c | c|}
\hline 
\multicolumn{2}{|c||}{\multirow{2}{*}{$A\rightarrow BC$}} & \multicolumn{2}{c|}{$B=\text{BSM}$} \\
\cline{3-4} 
\multicolumn{2}{|c||}{}  & 	$j j \ell $				&	$\ell \ell \ell$	 \\
\hline 
\hline 
\multirow{5}{*}{\rotatebox{90}{$C=\text{BSM}$}}& $Wb$, $tZ$,	&	 \multirow{2}{*}{ $LQ \to T^\prime N$ } 		&	 \multirow{2}{*}{ $LQ \to T^\prime L^\prime$ }	 \\
& $tH$, $tg$,  or $t \gamma$       &                                                                &                                                               \\
\cline{2-4}
&$Wt$, $bZ$,	&	 \multirow{2}{*}{  $LQ \to B^\prime N$ } 		&	 \multirow{2}{*}{  $LQ \to B^\prime L^\prime$} 	 \\
&$bH$, $bg$, or $b\gamma$      &                                                                &                                                               \\
\cline{2-4}
&$\ell Z$ or $\ell \gamma$	&	 $W^\prime \to L^\prime N$		&	$Z^\prime\to L^\prime L^\prime$	 \\
\hline
\end{tabular}
\caption{\label{tab:searchIX} Example (Group IX in Table \ref{tab:searches}) for $A \to B C$, where $A$, $B$ and $C$ are BSM particles.
The $C$ decays to two different SM particles, while the $B$ decays to more complex final states.
The $N$ denotes a right-handed heavy neutrino which decays to $N \to W^{\prime (*)} \ell \to j j \ell $.
The $L^\prime$ represents a generic vector-like lepton decaying to $L^\prime \to Z^{\prime (*)} \ell \to \ell \ell \ell$. 
The $T^\prime$, $B^\prime$, and $L^\prime$ decays are the same as presented in Table \ref{tab:searchV}.
}
\end{table}



\begin{table}[h!]
\centering
\renewcommand{\arraystretch}{1.2}
\scalebox{1}{
\begin{tabular}{|c|c|| c | c| c|}
\hline 
\multicolumn{2}{|c||}{\multirow{2}{*}{$A\rightarrow BC$}}& \multicolumn{3}{c|}{$B=\text{BSM}$} \\
\cline{3-5} 
\multicolumn{2}{|c||}{} 	& 	$N (\to j j \ell)$	&	$Q^\prime (\to jjj) $		& 	$LQ (\to \ell j)$ \\
\hline 
\hline 
\multirow{3}{*}{\rotatebox{90}{\hspace*{-0.1cm}$C=\text{BSM}$}}&$N (\to j j \ell)$	&	$Z^\prime \to N \bar N$		&	$Q^{\prime\prime} \to N \;Q^\prime$		&		\\
\cline{2-5} 
&$Q^\prime (\to jjj) $	&							& 	$Z^\prime \to Q^\prime \;\bar Q^\prime$	& 	$L^\prime \to Q^\prime \; LQ $ \\
\cline{2-5}
&$LQ (\to \ell j)$	&							&						& 	$Z^\prime \to LQ \; \overline{LQ} $ \\
\hline 
\end{tabular}
}
\caption{\label{tab:searchX} Example (Group X in Table \ref{tab:searches}) for $A \to B C$, where $A$, $B$ and $C$ are BSM particles.
Both $B$ and $C$ decay to more complex final states. 
The $N$ denotes a right-handed heavy neutrino which can decay into $N \to W^{\prime (*)} \ell \to j j \ell $.
The $Q^\prime$ represents a generic vector-like quark decaying to $Q^\prime \to W^{\prime (*)} j \to j j j$.} 
\end{table}


As an alternative example for X, we provide various coloron decays in Table \ref{tab:searchX2}. 
In this example, the $j$ includes a $t$, $b$, and light-flavor quarks.
The $G^\prime$, $\Theta$, and $\phi_I$ denote a coloron, color-octet scalar, and a singlet scalar respectively.
The three particles naturally arise in a `renormalizable coloron model' \cite{Bai:2018jsr, Bai:2018wnt}.
It is interesting that a simple coloron model provides such diverse signatures, depending on the mass spectrum.

\begin{table}[t!]
\centering
\renewcommand{\arraystretch}{1.1}
\scalebox{0.97}{
\begin{tabular}{|c|c|| c| c|c|c|c|c|c|c|c|}
\hline 
\multicolumn{2}{|c||}{\multirow{2}{*}{$A\rightarrow BC$}} & \multicolumn{7}{c|}{$B=\text{BSM}$} \\
\cline{3-9}
 \multicolumn{2}{|c||}{}  &   $gg jj$ & $gg jjjj$  & $WW jj$  & $\gamma Z jj$ & $Z Z jj$  & $H j j$  & $H jjjj$\\	
\hline 
\hline 
\multirow{5}{*}{\rotatebox{90}{\hspace*{-1cm}$C=\text{BSM}$}}& $ggjj$ 		 &   & $G^\prime \to\Theta \phi_I$ &  $G^\prime \to\Theta \phi_I$                      &    $G^\prime \to\Theta \phi_I$    & $G^\prime \to\Theta \phi_I$      &        & $G^\prime \to\Theta \phi_I$\\
\cline{2-9}
&$gg jjjj$	&                  & $G^\prime \to\Theta\Theta$ & $G^\prime \to\Theta\Theta$         &  $G^\prime \to\Theta\Theta$  &    $G^\prime \to\Theta\Theta$      &  $G^\prime \to\Theta \phi_I$     & $G^\prime \to\Theta\Theta$      \\
\cline{2-9}
&$WW jj$ 		 &                  &                          &   $G^\prime \to\Theta\Theta$ & $G^\prime \to\Theta\Theta$ & $G^\prime \to\Theta\Theta$    & $G^\prime \to\Theta \phi_I$   &  $G^\prime \to\Theta\Theta$ \\
\cline{2-9} 
&$\gamma Z j j$ 	 &                  &                          &             &      $G^\prime \to\Theta\Theta$   &   $G^\prime \to\Theta\Theta$     &  $G^\prime \to\Theta \phi_I$     &   $G^\prime \to\Theta\Theta$                   \\
\cline{2-9} 
&$Z Z j j$ 	 &                  &                          &             &      &   $G^\prime \to\Theta\Theta$       &  $G^\prime \to\Theta \phi_I$         &    $G^\prime \to\Theta\Theta$            \\
\cline{2-9}
&$H j  j$ 	 &                  &                          &             &      &          &          &   $G^\prime \to\Theta \phi_I$            \\
\cline{2-9} 
&$H j j j j$ 	 &                  &                          &             &      &          &           &   $G^\prime \to\Theta\Theta$            \\
\hline 
\end{tabular}
}
\caption{\label{tab:searchX2} Example (Group X in Table \ref{tab:searches}) for $A \to B C$, 
where $A$, $B$ and $C$ are BSM particles, where both $B$ and $C$ decay to more complex final states. 
The $j$ includes a $t$, $b$, and light-flavor quarks.
The $G^\prime$, $\Theta$, and $\phi_I$ denote a coloron, color-octet scalar, and a singlet scalar respectively \cite{Bai:2018jsr, Bai:2018wnt}.
So we could call all these entries as `a renormalizable coloron model'. It is interesting that a simple coloron model provides such diverse signatures, depending on the mass spectrum.}
\end{table}


Finally we make a brief remark on combining different groups. 
In Groups II and III, $A$ and $B$ are BSM resonances and $C$ is SM particle. 
Since $C$ is a SM particle, we can classify $A$ and $B$ based on the spin of $C$. 
Some examples are shown in Table \ref{tab:searchgroup1} for Groups II and III.
Any pair, $FF$, $FV$ etc only indicates Lorentz structure and they could have different (QED, QCD) charges.
$C$ could be $F$, $V$ or $H$, and $B$ (2-body resonance) will decay into any possible pair of $C$s. The spin of $A$ will be determined, once the spin of $B$ is chosen. All primed particles are BSM particles. In principle, this classification could include Group IV but would be more complicated. $X$ represents either $S$ or $V$.

\begin{table}[b!]
\centering
\begin{tabular}{|c|| c | c| c|c|c|c|}
\hline 
$A \to B C$	& 	$B\to FF$				& $B\to VV$				&	$B\to HH$	& $B\to FV$	& $B\to FH$	& $B\to VH$	 \\
\hline 
\hline 
C=F	&	$F^\prime\to F X^\prime$	& $F^\prime\to F X^\prime$	& $F^\prime\to F X^\prime$	& $X^\prime\to F F^\prime$	& $X^\prime\to F F^\prime$	& $F^\prime\to F X^\prime$ \\
\hline 
C=V	&	$V^{\prime\prime}\to V X^\prime$& $S^{\prime\prime}\to V X^\prime$	& $S^{\prime\prime}\to V X^\prime$	& $F^{\prime\prime}\to V F^\prime$	& $F^{\prime\prime}\to V F^\prime$	& $X^{\prime\prime}\to V X^\prime$ \\
\hline 
C=H	&	$X^{\prime\prime}\to H X^\prime$ & $X^{\prime\prime}\to H X^\prime$	& $X^{\prime\prime}\to H X^\prime$	&$F^{\prime\prime}\to H F^\prime$	& $F^{\prime\prime}\to H F^\prime$	& $X^{\prime\prime}\to H X^\prime$ \\
\hline 
\end{tabular}
\caption{\label{tab:searchgroup1} Example (Groups II and III in Table \ref{tab:searches}) for $A \to B C$ purely based on Lorentz structure, where $A$ and $B$ are BSM resonances and $C$ is SM particle. The corresponding (QED, QCD) charges need to be understood properly, depending on quantum charges of the involved particles. Any pair, $FF$, $FV$ etc only indicates Lorentz structure and they could have different (QED, QCD) charges.
$C$ could be $F$, $V$ or $H$, and $B$ (2 body resonance) will decay into any possible pair of $F$, $V$ or $H$. The spin of $A$ will be determined, once the spin of $B$ is chosen. All primed particles are BSM particles. In principle, this classification could include III but would be more complicated. $X$ is either a scalar ($S$) or a vector ($V$).}
\end{table}


Similarly we can combine Groups V, VI and VIII, and show generic presentation of Lorentz structure in Table \ref{tab:searchgroup2}.
Here $A$, $B$ and $C$ are BSM resonances and both $B$ and $C$ could decay into any possible pair of $F$, $V$ or $H$. The spin of $A$ will determined, once the spins of $B$ and $C$ are chosen. In principle, this classification could include Groups VII, IX and X but would be more complicated.
The point of this exercise in Tables \ref{tab:searchgroup1} and \ref{tab:searchgroup2} is that we can find an example of any resonance, once we specify (QED, QCD) quantum numbers and Lorentz structure.  

\begin{table}[t!]
\centering
\scalebox{0.93}{
\begin{tabular}{|c|| c | c| c|c|c|c|}
\hline 
$A\to B C$		& 	$B\to FF$				& $B\to VV$				&	$B\to HH$	& $B\to FV$	& $B\to FH$	& $B\to VH$	 \\
\hline 
\hline 
$C\to FF$	& $X^{\prime\prime\prime} \to X^{\prime\prime} X^{\prime}$ 	& $X^{\prime\prime\prime} \to X^{\prime\prime} X^{\prime}$	& $X^{\prime\prime\prime} \to X^{\prime\prime} X^{\prime}$	& $F^{\prime\prime} \to X^\prime F^\prime$	& $F^{\prime\prime} \to X^\prime F^\prime$	&$X^{\prime\prime\prime} \to X^{\prime\prime} X^{\prime}$	\\
\hline 
$C\to VV$	& 	& $X^{\prime\prime\prime} \to X^{\prime\prime} X^{\prime}$	& $X^{\prime\prime\prime} \to X^{\prime\prime} X^{\prime}$	& $F^{\prime\prime} \to X^\prime F^\prime$	& $F^{\prime\prime} \to X^\prime F^\prime$	&$X^{\prime\prime\prime} \to X^{\prime\prime} X^{\prime}$	\\	
\hline 
$C\to HH$	&  	&	& $X^{\prime\prime\prime} \to X^{\prime\prime} X^{\prime}$	& $F^{\prime\prime} \to X^\prime F^\prime$ 	& $F^{\prime\prime} \to X^\prime F^\prime$	&	$X^{\prime\prime\prime} \to X^{\prime\prime} X^{\prime}$\\
\hline 
$C\to FV$	&  	&	&	& $X^{\prime} \to F^\prime F^{\prime\prime}$	& $X^{\prime} \to F^\prime F^{\prime\prime}$	& $F^{\prime\prime} \to F^\prime X^{\prime}$	\\
\hline 
$C\to FH$	&  	&	&	&	&$X^{\prime\prime\prime} \to X^{\prime\prime} X^{\prime}$	& $F^{\prime\prime} \to F^\prime X^{\prime}$	\\
\hline 
$C\to VH$	& 	&	&	&	&	& $X^{\prime\prime\prime} \to X^{\prime\prime} X^{\prime}$	\\
\hline 
\end{tabular}
}
\caption{\label{tab:searchgroup2} Example (Groups V, VI and VIII in Table \ref{tab:searches}) for $A \to B C$ purely based on Lorentz structure, where $A$, $B$ and $C$ are BSM resonances. The corresponding (QED, QCD) charges need to be understood properly, depending on quantum charges of the involved particles. Any pair, $FF$, $FV$ etc only indicates Lorentz structure and they could have different QED/QCD charges.
Both $B$ and $C$ could decay into any possible pair of $F$, $V$ or $H$. The spin of $A$ will determined, once the spins of $B$ and $C$ are chosen. In principle, this classification could include VII, IX and X.}
\end{table}


\section{Current Status}
\label{sec:results}

ATLAS and CMS have an impressive and extensive search program that already includes many of the possibilities described in the previous sections.  In particular, a few more of the $A\rightarrow\text{SM}\times\text{SM}$ possibilities described in Ref.~\cite{Craig:2016rqv} are now covered by Run 2 searches.  Table~\ref{tab:searchesexp} describes the current coverage to both the $\text{SM}\times\text{SM}$ and more generic 2-body resonances cases using published searches based on Run 2 data.

The first important feature of Table~\ref{tab:searchesexp} is that many of the $\text{SM}\times\text{SM}$ possibilities are still uncovered, most notably the final states involving a lepton and a quark/gluon or Higgs boson.
The second important feature of Table~\ref{tab:searchesexp} is that when one or both of $B/C$ are BSM, most of the possibilities are uncovered.   In some cases, such as $B/C\rightarrow$ quarks/gluons, there is some complementarity with direct $B/C$ searches (see Section~\ref{sec:motivation}).  This is also true when $B$ or $C$ decay into leptons or vector bosons, but the $B/C$ search limits are much weaker due to the low production cross-section of vector boson fusion at the LHC and the available center-of-mass energy at current and previous lepton colliders.

Despite a large number of existing searches, Table~\ref{tab:searchesexp} combined with Section~\ref{sec:motivation} indicate that there are many well-motivated possibilities that are currently uncovered.  New searches can close these gaps in coverage and ensure broad sensitivity to BSM possibilities. 

\begin{table}[h!]
\renewcommand{\arraystretch}{1.3}
\begin{adjustbox}{width=\columnwidth,center}

    \begin{tabular}{|cc||ccccccccc|cccc|cccc|cccc|}
    \hline
       && \multirow{2}{*}{$e$} &  \multirow{2}{*}{$\mu$} &  \multirow{2}{*}{$\tau$} &  \multirow{2}{*}{$q/g$} &  \multirow{2}{*}{$b$} &  \multirow{2}{*}{$t$} &  \multirow{2}{*}{$\gamma$} &  \multirow{2}{*}{$Z/W$} &  \multirow{2}{*}{$H$} & \multicolumn{4}{|c|}{$\text{BSM}\rightarrow \text{SM$_1$}\times\text{SM$_1$}$} &  \multicolumn{4}{|c|}{$\text{BSM}\rightarrow \text{SM$_1$}\times\text{SM$_2$}$} & \multicolumn{4}{|c|}{$\text{BSM}\rightarrow \text{complex}$}\\
       & & & & & & & & & &  & $q/g$ & $\gamma/\pi^0$'s & $b$ &$  \cdots$ & $tZ/H$ & $bH$ & &$  \cdots$ & $\tau qq'$ & $eqq'$ & $\mu qq'$ & $\cdots $ \\
       \hline
       \hline
      \multicolumn{2}{|c||}{$e$} &\cellcolor{blue!10}\cite{ATLAS-CONF-2019-001,Sirunyan:2018exx}  &\cellcolor{blue!10} \cite{Aaboud:2018jff,Sirunyan:2018zhy}&\cellcolor{blue!10}\cite{Aaboud:2018jff}&$\emptyset$&$\emptyset$&$\emptyset$&\cite{Sirunyan:2018zzr}&\cite{Sirunyan:2018mtv}&$\emptyset$&$\emptyset$&$\emptyset$&$\emptyset$&&$\emptyset$&$\emptyset$&$\emptyset$&&$\emptyset$&\cite{Sirunyan:2018pom,Sirunyan:2017xnz}&$\emptyset$&\\
       \multicolumn{2}{|c||}{$\mu$}& &\cellcolor{blue!10}\cite{ATLAS-CONF-2019-001,Sirunyan:2018exx}&\cellcolor{blue!10}\cite{Aaboud:2018jff}&$\emptyset$&$\emptyset$&$\emptyset$&\cite{Sirunyan:2018zzr}&\cite{Sirunyan:2018mtv}&$\emptyset$&$\emptyset$&$\emptyset$&$\emptyset$ &&$\emptyset$&$\emptyset$&$\emptyset$& &$\emptyset$&$\emptyset$&\cite{Sirunyan:2018pom,Sirunyan:2017xnz}&\\
       \multicolumn{2}{|c||}{$\tau$}& & &\cellcolor{blue!10}\cite{Aaboud:2017sjh,Khachatryan:2016qkc}&$\emptyset$&\cite{Sirunyan:2018jdk}&$\emptyset$&$\emptyset$&$\emptyset$&$\emptyset$&$\emptyset$&$\emptyset$& $\emptyset$&&$\emptyset$&$\emptyset$&$\emptyset$& &\cite{Sirunyan:2018vhk,Sirunyan:2017yrk}&$\emptyset$&$\emptyset$&\\
        \multicolumn{2}{|c||}{$q/g$} & & &&\cellcolor{blue!10}\cite{Aaboud:2019zxd,ATLAS:2015nsi,Sirunyan:2018xlo,Sirunyan:2017nvi}&\cellcolor{blue!10}\cite{Aaboud:2018tqo}&$\emptyset$&\cellcolor{blue!10}\cite{Aaboud:2017nak,Sirunyan:2017fho}&\cellcolor{blue!10}\cite{Sirunyan:2017acf}&$\emptyset$&$\emptyset$&$\emptyset$& $\emptyset$&&$\emptyset$&$\emptyset$&$\emptyset$& &$\emptyset$&$\emptyset$&$\emptyset$&\\
          \multicolumn{2}{|c||}{$b$}& &&&&\cellcolor{blue!10}\cite{Aaboud:2019zxd,Aaboud:2018tqo,Sirunyan:2018ikr}&\cellcolor{blue!10}\cite{Aaboud:2018jux}&\cite{Sirunyan:2017fho}&\cite{Aaboud:2018ifs}&\cite{ATLAS-CONF-2018-024}&$\emptyset$&$\emptyset$& $\emptyset$&&\cite{Sirunyan:2018fki}&$\emptyset$&$\emptyset$& &$\emptyset$&$\emptyset$&$\emptyset$& \\
           \multicolumn{2}{|c||}{$t$} & &&&&&\cellcolor{blue!10}\cite{Aaboud:2019roo}&$\emptyset$&\cellcolor{blue!10}\cite{Sirunyan:2019tib}&\cite{Sirunyan:2016ipo}&$\emptyset$&$\emptyset$&$\emptyset$ &&\cite{Sirunyan:2018rfo}&\cite{Sirunyan:2018fki}&$\emptyset$& &$\emptyset$&$\emptyset$&$\emptyset$& \\
           \multicolumn{2}{|c||}{$\gamma$} & & &&&&&\cellcolor{blue!10}\cite{Aaboud:2017yyg,Sirunyan:2018wnk}&\cellcolor{blue!10}\cite{Aaboud:2016trl,Aaboud:2018fgi,Sirunyan:2017hsb}&\cite{Aaboud:2018fgi,Sirunyan:2018dsh}&$\emptyset$&$\emptyset$& $\emptyset$&&$\emptyset$&$\emptyset$&$\emptyset$& &$\emptyset$&$\emptyset$&$\emptyset$&\\
            \multicolumn{2}{|c||}{$Z/W$} & & &&&&&&\cellcolor{blue!10}\cite{Aaboud:2018bun}&\cellcolor{blue!10}\cite{Aaboud:2018bun}&$\emptyset$&$\emptyset$& $\emptyset$&&$\emptyset$&$\emptyset$&$\emptyset$& &$\emptyset$&$\emptyset$&$\emptyset$&\\
            \multicolumn{2}{|c||}{$H$} & & &&&&&&&\cellcolor{blue!10}\cite{Aaboud:2018sfw,Aaboud:2018knk}&\cite{Aaboud:2017ecz}&$\emptyset$& $\emptyset$&&$\emptyset$&$\emptyset$&$\emptyset$& &$\emptyset$&$\emptyset$&$\emptyset$&\\
             \hline
           \multirow{2}{*}{ \rotatebox{90}{$\text{BSM}\rightarrow \text{SM$_1$}\times\text{SM$_1$~~~}$}}& $q/g$&&&&&&&&&&$\emptyset$& $\emptyset$&$\emptyset$&&$\emptyset$&$\emptyset$&$\emptyset$ &&$\emptyset$&$\emptyset$&$\emptyset$& \\ 
           & $\gamma/\pi^0$'s & &&&&&&&&&&\cite{Aaboud:2018djx} &$\emptyset$&&$\emptyset$&$\emptyset$&$\emptyset$ &&$\emptyset$&$\emptyset$&$\emptyset$&\\
            & $b$ &&&&&&&&&&& &\cite{Aaboud:2018aqj,Aaboud:2019opc}&&$\emptyset$&$\emptyset$& $\emptyset$&&$\emptyset$&$\emptyset$&$\emptyset$&\\
             & \vdots &&&&&&&&&& &&&&& &&&&&&\\
              &  &&&&&&&&&&& &&&&& &&&&&\\
             &  &&&&&&&&&&& &&&&& &&&&&\\
                  & \vdots &&&&&&&&&& &&&&& &&&&&&\\
                       \hline
           \multirow{2}{*}{ \rotatebox{90}{$\text{BSM}\rightarrow \text{SM$_1$}\times\text{SM$_2$ }~~~$}}&  $tZ/H$&&&&&&&&&&& &&&&& &&&&& \\ 
           & $bH$&&&&&&&&&&& &&&&& &&&&&\\
            & \vdots &&&&&&&&&&& &&&&& &&&&&\\
             &  &&&&&&&&&&& &&&&& &&&&&\\
              &  &&&&&&&&&&& &&&&& &&&&&\\
                  &  &&&&&&&&&&& &&&&& &&&&&\\
                      & \vdots &&&&&&&&&&& &&&&& &&&&&\\    
                       \hline 
             \multirow{2}{*}{ \rotatebox{90}{$\text{BSM}\rightarrow\text{complex }~~~$}}& $\tau qq'$& &&&&&&&&&& &&&&& &&&&&\\ 
           & $eqq'$ &&&&&&&&&&& &&&&& &&&&&\\
            & $\mu qq'$ &&&&&&&&&&& &&&&& &&&&&\\
             & \vdots &&&&&&&&&&& &&&&& &&&&&\\
              &  &&&&&&&&&&& &&&&& &&&&&\\
                      & \vdots &&&&&&&&&&& &&&&& &&&&&\\      
                       \hline           
    \end{tabular}

  \end{adjustbox}

  \caption{\label{tab:searchesexp}References to existing searches for two-body resonances, where one decay product is from the first column and one is from the first row.  Only the most recent searches are considered.  The box $\text{BSM}\rightarrow \text{SM$_1$}\times\text{SM$_2$ }$ represents cases where the primary resonance decays to a BSM particle, which itself decays into two SM particles that are not the same.  \colorbox{blue!20}{Colored cells} indicate searches that were covered by $\sqrt{s}=8$ TeV searches reported in Ref.~\cite{Craig:2016rqv}.}
\end{table}


\section{Conclusions}
\label{sec:conclusions}

Two-body resonance searches are a cornerstone of the LHC search program.  While the current experimental coverage is broad, there are many well-motivated scenarios that are all or partially uncovered.  We have catalogued the set of possibilities, providing at least one motivating example for each final state.  Given the lack of significant excess at the LHC and the lack of a unique theory to guide the search program, now is the time to consider diversifying the experimental sensitivity.  Organizing the possibilities by final state provides a way forward.

While the traditional search program will be able to accommodate many of the possibilities described earlier, there are not enough resources to consider all potential final states.  Therefore, dedicated searches will likely need to be complimented with more model agnostic searches.  Machine learning methods may be able to automate this approach and solve significant statistical challenges like large trails factors~\cite{Collins:2019jip,Collins:2018epr}.  In particular, techniques such as neural networks can readily analyze high-dimensional spaces and approaches with cross-validation can avoid over-training.

This work has focused on two-body decays into visible final states.  Future work will consider cases where there are undetectable particles (such as neutrinos and dark sectors) as well as multi-body decays.

The LHC experiments have and will continue to collect rich datasets that may contain answers to key questions about the fundamental properties of nature.  Many well-motivated fundamental theories have provided guiding principles to analyses these data.  However, a more diversified perspective will be required to full exploit the data - in fact, there may be something new already hiding in the existing datasets!

\section{Acknowledgments}

This work was supported by the U.S.~Department of Energy, Office of Science under contract DE-AC02-05CH11231, DE-SC0017988 and DE-SC0019474.

\bibliographystyle{elsarticle-num}
\bibliography{myrefs}{}

\end{document}